\pdfoutput=1
\documentclass[12pt]{article}
\usepackage{a4wide}
\usepackage{latexsym}
\usepackage{epsfig}
\usepackage{amsmath}
\usepackage{amssymb}
\usepackage{url}
\usepackage{subfigure}
\usepackage{cancel}

\newlength{\dinwidth} \newlength{\dinmargin}
\begin{document}

\thispagestyle{empty}

\begin{flushright}
  Nikhef-2010-011\\
  ITP-UU-10/16\\
  ITFA-2010-14\\
\end{flushright}

\vspace{1.5cm}

\begin{center}
 {\Large\bf Single top production in a non-minimal supersymmetric model}\\[1cm] 
  {\sc Michel Herquet$^{a}$, Robert Knegjens$^{a}$, Eric Laenen$^{a,b,c}$\\
  \vspace{1.5cm}
  $^a${\it Nikhef Theory Group\\
    Science Park 105, 1098 XG Amsterdam, The Netherlands} \\[0.4cm]
  $^b${\it Institute for Theoretical Physics, Utrecht University\\
    Leuvenlaan 4, 3584 CE Utrecht, The Netherlands} \\[0.4cm]
  $^c${\it Institute for Theoretical Physics, University of Amsterdam\\
    Valckenierstraat 65, 1018 XE Amsterdam, The Netherlands} \\    
}
\end{center}

\vspace{2cm}

% ### abstract
% ####################################################################

\begin{abstract}

\noindent{
We study single top production at the LHC in a SUSY-QCD model with a heavy Dirac gluino.
The presence of a heavy Dirac gluino allows for notable top-up flavour changing neutral currents. 
In this scenario, we find that the process $ug\to tg$ gives the largest contribution to single top production via FCNCs at the LHC. 
The key features of this signal are that the top quark is produced very forward and that it is asymmetric to its anti-top counterpart, as the latter lacks a valence quark.
}

\end{abstract}

\vspace*{\fill}

\newpage
\reversemarginpar

% ### Introduction
% ####################################################################
\section{Introduction}
\label{sec:intro}

Due to its large mass, the top quark's presence as an initial partonic state at hadron colliders is negligible.
The single production of top quarks must, therefore, proceed via flavour changing interactions.
In the Standard Model (SM) the tree level couplings of the $W$ bosons to quarks generate all such interactions.
At the loop level, flavour changing neutral current (FCNC) interactions are possible.
Single top production via FCNCs in the SM is, however, strongly suppressed by the Glashow-Iliopoulos-Maiani (GIM) mechanism due to the small mass differences of the down type quarks occurring in the loop.

In supersymmetric extensions of the SM, flavour mixing in the squark sector allows for additional top quark FCNCs.
Unlike the down type quarks, the mediating squarks may have suitably different masses to avoid a similar supression from a GIM-like mechanism. 
In the Minimal Supersymmetric Standard Model (MSSM), however, squark flavour mixing is strongly constrained due to the excellent agreement between the SM and experiment~\cite{Gabbiani:1988rb,Hagelin:1992tc,Gabbiani:1996hi, Misiak:1997ei,Altmannshofer:2009ne}. 
The flavour mixing of the first generation of squarks with those of the second and third are constrained by the $K^0$, $B^0_d$ and $D^0$ neutral meson mixing experiments.
Likewise, the flavour mixing of the second and third generations of squarks is constrained by $B^0_s$ mixing as well as experimental results from the FCNC process $b \to s\gamma$.

Phenomenological studies of top quark FCNC processes in the MSSM have predominantly focused on the less-constrained stop-scharm mixing~\cite{Liu:2004bb,Eilam:2006rb,Guasch:2006hf,LopezVal:2007rc,Cao:2006xb,Cao:2007dk,Plehn:2009it}.
The dominant contributions are found to come from the SUSY-QCD sector, namely, from gluino, as opposed to neutralino, exchange.
For a review of top quark FCNCs in new physics and a more detailed list of references see, for example, Ref.~\cite{Larios:2006pb}.

Recently, it has been shown that sizable squark flavour mixing is possible in models with heavy Dirac gauginos~\cite{Kribs:2007ac,Blechman:2008gu}. 
Specifically, it was found that heavy Dirac gauginos suppress the SUSY contributions to neutral meson mixing, thereby relaxing the associated experimental constraints on squark mixing.
The Dirac gauginos were considered in the context of the Minimal R-symmetric Supersymmetric Standard Model (MRSSM): a SUSY extension of the SM in which the global $U(1)_R$ symmetry of the N=1 superalgebra remains unbroken.
In such models gauginos are required to be Dirac fermions because their charge under the continuous R symmetry forbids a Majorana mass term.
One means of promoting gauginos to Dirac fermions involves the addition of adjoint chiral superfields together with a supersoft breaking mechanism~\cite{Fox:2002bu}.
Alternatively, Dirac gauginos appear automatically in the hypermultiplets of N=2 SUSY.

In this paper we study the effects of the extra squark mixing freedom that heavy Dirac gluinos allow on single top production.
In this context the squark mixing of the third generation with the first is more relevant than with the second (stop-scharm), as only the former has all its constraints relaxed.
Single top production has already been considered in the MRSSM via the process of tree level squark pair production, with the squarks decaying to top or lighter quarks and gauginos~\cite{Kribs:2009zy}.
Our focus will be on single top production via top-up FCNC processes in SUSY QCD.

% ### Single top in SUSY-QCD
% ####################################################################
\section{Single top in SUSY-QCD}
\label{sec:obs}

% ### Top quark FCNC couplings
% ###--------------------------------------------------------------###
\subsection{Top quark FCNC couplings}
\label{subsec:top_fcnc}

The additional quark flavour mixing possible in SUSY follows from a misalignment between the squark and quark mass eigenstates.
Taking the quark mass matrix to be diagonal, the mixing is encoded in the off-diagonal elements of the squark mass matrix.
Contributions to the squark mass matrix come from the soft breaking mass terms as well as the superpotential, $D$-terms and soft breaking trilinear terms after electroweak symmetry breaking.
Mixing between left and right (LR) labeled squarks is only possible through the electroweak breaking of the superpotential $\mu$ term and the soft trilinear terms. 
As both of these terms violate a $U(1)$ $R$ symmetry, the MRSSM cannot have LR squark mixing.
Consequently, LL and RR squark mixing can be larger within the total mixing limits. 
To remain consistent with the analysis of Ref.~\cite{Kribs:2007ac} we will consider only LL and RR mixing.
The squark mixing matrix is then given by
\begin{align}
  \label{eq:squarkmass}
  \left(\mathcal{M}_{\tilde{q}}^2\right)_{ij} =& \left(\tilde{m}_q^2\right)_{ij}
  + \Big[m_{q_3}^2 + M_z^2 \cos(2\beta)\left(T_3 - Q\sin^2\theta_W\right)\Big] \delta_{i3}\delta_{j3},
\end{align}
with $q \in\{u_L, d_L, u_R, d_R \}$, $m_{u_3} = m_t$, $m_{d_3} = m_b$ and $i$ is the generation index of the quark mass eigenstate basis. 
The matrices $\tilde{m}_q^2$ are soft SUSY breaking mass terms, which by SU(2) symmetry obey the relation $\tilde{m}_{uL}^2 = \tilde{m}_{dL}^2$.
The mass eigenstates of the squarks can be found by diagonalizing the above mass matrix:
\begin{equation}
  \label{eq:diagonalize}
  -\tilde{q}_i^* \left(\mathcal{M}_{\tilde{q}}^2\right)_{ij} \tilde{q}_j 
  = - \tilde{q}_a^* \left( U^\dagger_{\tilde{q}} \mathcal{M}^2_{\tilde{q}} U_{\tilde{q}}\right)_{ab} \tilde{q}_b,
\end{equation}
with
\begin{equation}
  \label{eq:squarkmix}
  \tilde{q}_a = \sum_i \left( U_{\tilde{q}}^\dagger\right)_{ai} \tilde{q}_i.
\end{equation}
Here $U_{\tilde{q}}$ is a unitary matrix and the index $a$ denotes the squark mass eigenstate basis.

\begin{figure}[htbp]
  \centering
  \includegraphics[width=130mm]{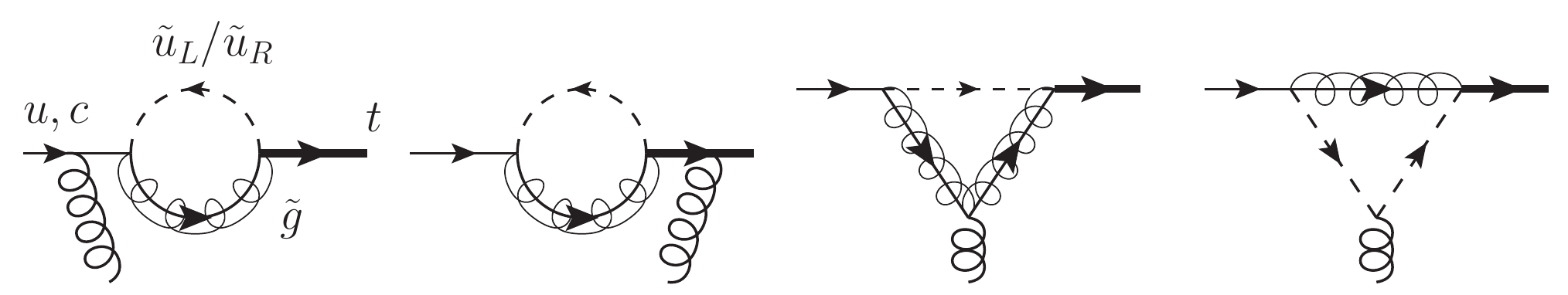}
  \caption{The bubble and triangle diagrams contributing to the $t$-$u$-$g$ FCNC. The arrows on the gluino lines are indicative of their Dirac fermion nature. \label{fig:topFCNCs}}
\end{figure}

The dominant top quark FCNC couplings in SUSY QCD are given by the one-loop bubble and triangle diagrams shown in Figure~\ref{fig:topFCNCs}.
The flavour mixing matrices $U_{\tilde{q}}$ enter through the standard quark-gluino-squark vertices that are also present in the MSSM. Specifically, via the Feynman rules:
\begin{align}
\bar{q}_{i} \text{ - } \tilde{g}_A\text{ - }\tilde{q}^*_{\sigma a}:&\quad -i \kappa_\sigma \sqrt{2} g_S \big(U^\dagger_{\tilde{q}\sigma} \big)_{ai} t_A P_{\sigma},\notag\\
q_{i}\text{ - }\overline{\tilde{g}}_A\text{ - }\tilde{q}_{\sigma a}:&\quad -i \kappa_\sigma \sqrt{2} g_S \big(U_{\tilde{q}\sigma} \big)_{ai} t_A P_{\overline{\sigma}}, 
\end{align}
with $\sigma \in\{L, R\}$, $\kappa_\sigma = \{+1:L,-1:R\}$ and $\bar{\sigma}= \sigma(L \leftrightarrow R)$.
The amplitudes of these diagrams are vulnerable to a GIM-like suppression, as they each involve the sum over orthogonal elements of the unitary matrix $U_{\tilde{u}}$:
\begin{equation}
  \label{eq:gimsuppr}
  \mathcal{A} \propto \sum_{\sigma\in\{L,R\}} \sum_a \big( U_{\tilde{u}\sigma} \big)_{ia} 
  \big( U_{\tilde{u}\sigma}^\dagger \big)_{a3} f(\sigma,a),
\end{equation}
for $i\in\{1,2\}$, where the function $f$ represents the amplitudes dependence on the up squark masses. If the up squark masses are degenerate the amplitude will vanish. 
In general, the greater the mass difference (or splitting) in the squark sector, the larger this amplitude will be.
Note that the UV divergent terms of these loop diagrams vanish by this mechanism.

% ### Single top with Dirac gluinos 
% ###--------------------------------------------------------------###
\subsection{Single top with Dirac gluinos}
\label{subsec:dirac_gluinos}

As discussed in the introduction, the presence of Dirac as opposed to Majorana gluinos can suppress the SUSY-QCD box diagrams that contribute to neutral meson mixing phenomenology.
The requirement for this suppression to occur is that the gluino, besides from being Dirac, is sufficiently heavier than the squarks in the loop.
If this is the case, the strong constraints placed on first generation squark flavour mixing by meson mixing experiments are relaxed.
In this paper we are concerned with single top production and are thus only interested in the mixing between the first and third squark generations.
For simplicity, we parametrize the unitary matrices $U_{\tilde{q}}$ by a single Euler angle $\theta_q$ for each $q \in\{u_L, d_L, u_R, d_R \}$.

\begin{figure}[htbp]
  \centering
  \includegraphics[width=130mm]{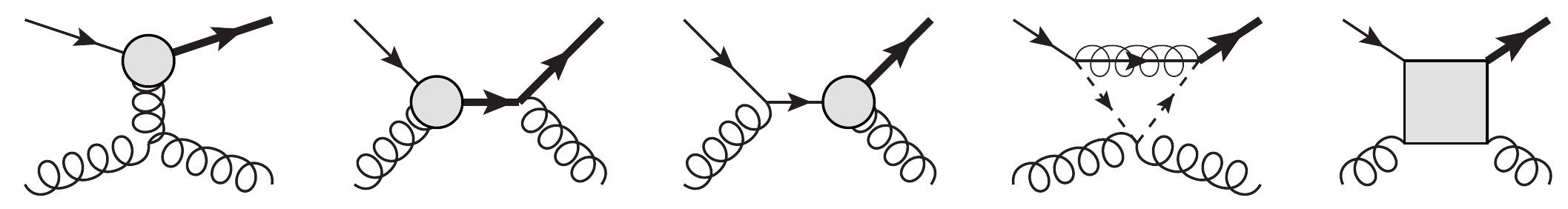}
  \caption{Feynman diagrams contributing to $ug\to tg$. The shaded $t$-$u$-$g$ vertex represents the diagrams in Figure~\ref{fig:topFCNCs}. Cross-channel gluon diagrams are not shown.  \label{fig:diagrams}}
\end{figure}

The production of a single top quark in the SM at tree level is accompanied by an additional lighter parton that eventually hadronizes into a jet.
With the top-up FCNC discussed above, it is possible to produce exclusively single top quarks without this extra jet~\cite{Plehn:2009it}.
Our focus will nonetheless be on the production of a single top plus jet in SUSY-QCD, as it mimics the SM single top signal that is about to be extensively scrutinized at the upcoming LHC.
Top-up FCNC processes that give single top plus jet are $ug\to tg$, $gg\to t\bar{u}$, $uq\to tq$, $u\bar{q}\to t\bar{q}$ and $q\bar{q}\to t\bar{u}$ where $q\in \{u,d,s,c,b\}$.
In Figure~\ref{fig:diagrams} we present the diagrams that contribute to the process $ug\to tg$.

The relevant parameters for the SUSY-QCD FCNC couplings are the gluino mass $m_{\tilde{g}}$, the squark mixing angles $\theta_{uL}$, $\theta_{uR}$ and the 1st and 3rd generation squark masses $m_{\tilde{u}L a},$ $m_{\tilde{u}R a}$.
We make no attempt to calculate what regions of the parameter space satisfy the shifted mixing constraints beyond what was done by Ref.~\cite{Kribs:2007ac}.
Rather, we take a pragmatic approach by assigning a prototype point in parameter space and in turn vary the parameters with respect to it.
We take this point to have a heavy gluino $m_{\tilde{g}} = 2$~TeV, maximal squark mixing $\theta_{uL}=\theta_{uR} =\pi/4$ and squark masses with a sizable mass splitting $m_{\tilde{1}L} = 400$~GeV, $m_{\tilde{3}L} = 1000$~GeV.
Right handed squark masses are set to 90\% of their left handed counterparts.

% ### Signals at the LHC 
% ####################################################################
\section{LHC Phenomenology}
\label{sec:pheno}

% ### Signal Features 
% ###--------------------------------------------------------------###
\subsection{Signal Features}
\label{subsec:asymmetry}

To calculate the partonic one loop squared amplitudes giving top plus jet we made use of the software packages FeynArts, FormCalc and LoopTools~\cite{Hahn:1998yk,Hahn:2000kx,Hahn:2009bf,Vermaseren:2008kw}.
The LHC hadronic cross sections were computed using the CTEQ6M PDF sets~\cite{Pumplin:2002vw}, with the renormalization and factorization scales set to the top quark mass and the LHC centre of mass energy taken to be 14 TeV.
To stay within the jet detection limits at the LHC, a maximum pseudorapidity cut of 3 and a minimum transverse momentum cut of 20 GeV are placed on the outgoing parton accompanying the top quark. 

Of the single top plus jet processes possible via the top-up FCNC couplings, we found the process $ug\to tg$ to be by far the most dominant at the LHC with a contribution of more than 80\%.
This can be attributed to the up quark density in the colliding protons, together with the exchange of t-channel gluons within its leading diagrams.
At the prototype parameter point discussed in the previous section, $ug\to tg$ has a hadronic cross section of 8.5 pb.
This is in contrast to the runner-up process $uu\to tu$, which has a hadronic cross section of 404 fb. 
A model independent study of a $t$-$u$-$g$ chromo-magnetic coupling reports similar results~\cite{Han:1998tp}.
The dominance of $ug\to tg$ could give an observable and characteristic asymmetry between single top and anti-top production at the LHC, as the process $\bar{u}g \to \bar{t} g$ does not involve valence quarks.

\begin{figure}[htbp]
  \centering
  \subfigure[] {
      \label{fig:forwardTop}
      \includegraphics[width=7.5cm]{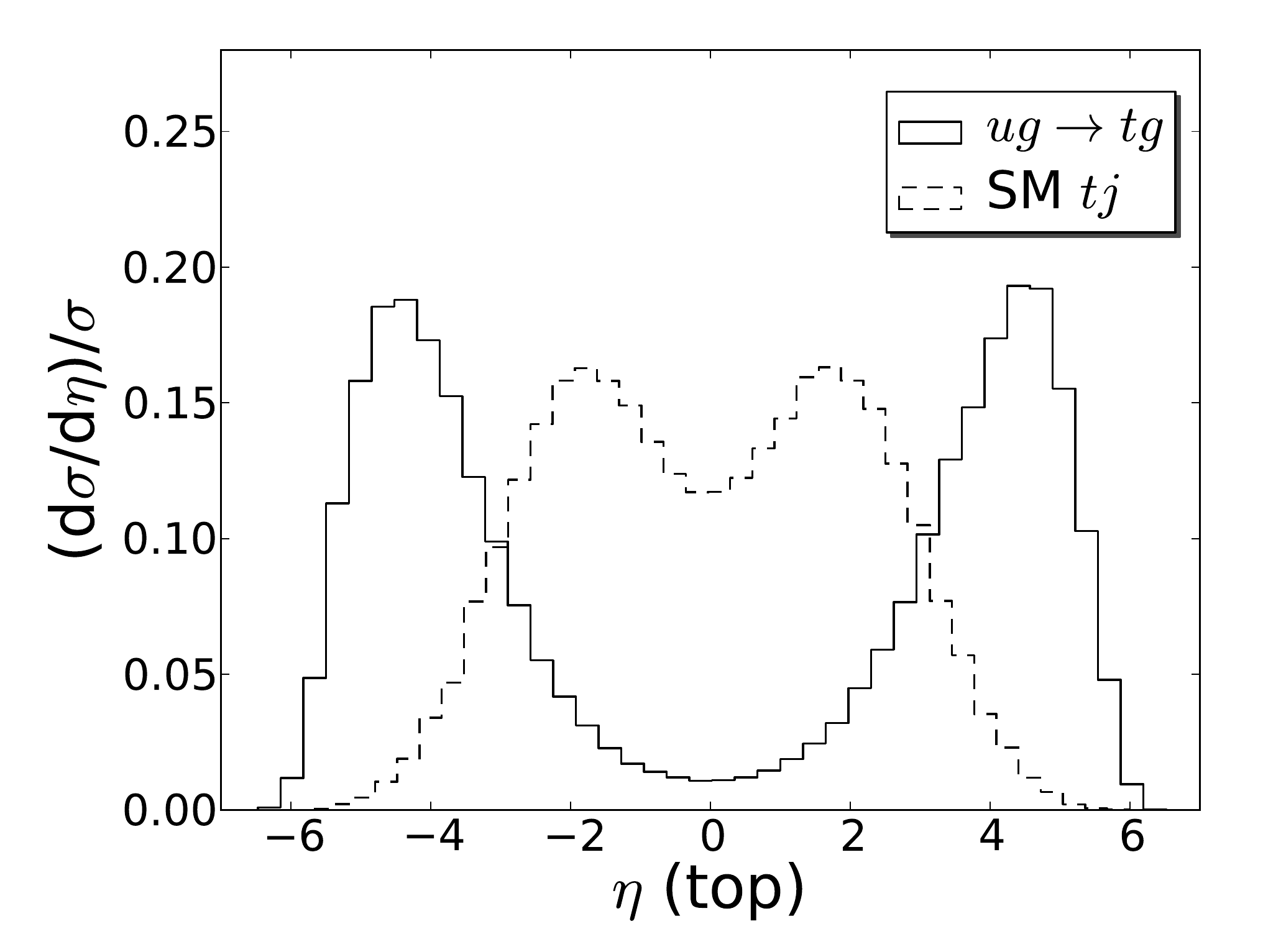}
  }
  \subfigure[] {
      \label{fig:forwardLepton}
      \includegraphics[width=7.5cm]{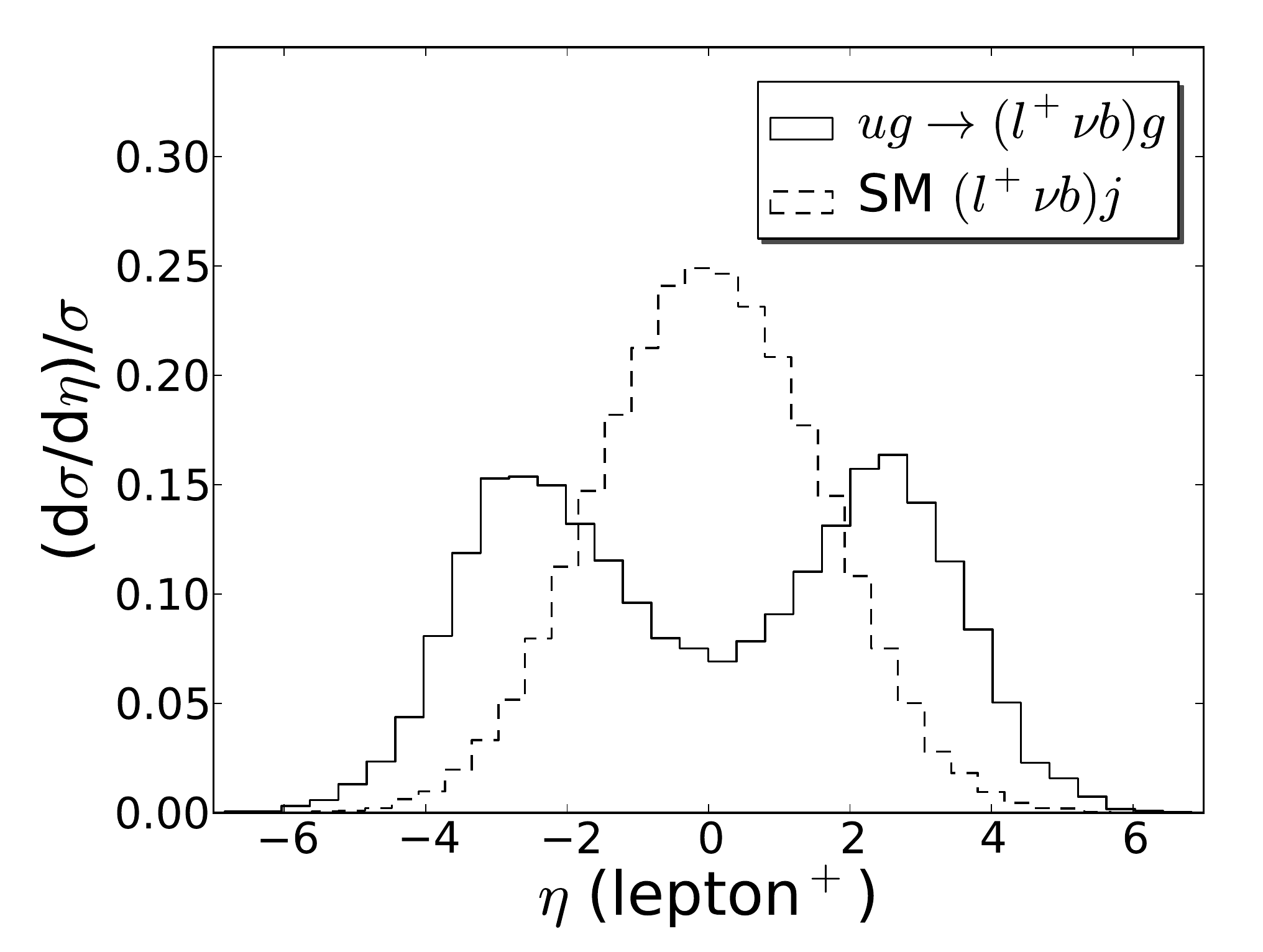}
  }
  \caption{The FCNC process $ug\to tg$, as well its semileptonic decay, have a distinct forward signature to their SM counterparts at the LHC. Figures (a) and (b) give the normalized pseudorapidity distributions for the top quark and charged lepton respectively. Only the parton accompanying the top has been cut on with $\eta$~$<$~3 and $p_T$~$>$~20~GeV }
\end{figure}

A key feature of the $ug\to tg$ process is that the top quark is produced very forward, as shown in Figure~\ref{fig:forwardTop} .
This makes it distinct from SM single top, for which the differential cross section is peaked more centrally.
For the semi-leptonic top decay this forwardness is carried over to the charged lepton, as shown in Figure~\ref{fig:forwardLepton}.
The detection of this signal at the LHC is therefore reliant on the forward efficiency of the detector. 

The one loop process $ug\to tg$ is dominated by the bubble and triangle $t$-$u$-$g$ couplings shown in Figure~\ref{fig:topFCNCs}.
A useful feature of these couplings is that their leading form factors have limited kinematic dependence and therefore vary little across the LHC phase space.
This was tested numerically using FormCalc, with the triangle coupling taken to be on-shell to reduce the number of form factors from 16 to four.
The triangle and bubble form factors that contribute most significantly are those involving the spinor structure $\gamma^\mu P_{L/R}$ and $P_{L/R}$ respectively.
It is thus possible to approximate the signal process by using constant tree level couplings with this spinor structure.

% ### Signal versus background
% ###--------------------------------------------------------------###
\subsection{Signal and background}
\label{subsec:signal}

To assess the strength of our signal over the SM background at the LHC we used MadGraph/MadEvent to generate events~\cite{Maltoni:2002qb,Alwall:2007st}.
For the signal events an effective tree level $t$-$u$-$g$ vertex with a constant form factor was used, as discussed in the previous section.
Normalized against the full cross section as computed by FormCalc, the effective vertex was found to accurately reproduce the original kinematic distributions.
Subsequently, the Madgraph DECAY package was used to decay the top plus jet signal to a charged lepton, neutrino, b-jet and jet.
% *-- NEW:
To simulate detector smearing effects, the four momenta of the final state quarks and gluons are scaled by a Gaussian distribution that is centred around their energy.
% --*
The model parameters are again set to the prototype point discussed in Section~\ref{subsec:dirac_gluinos}.

% *-- MODIFIED:
%The acceptance cuts for the signal and background are taken to be $\eta$~$<$~5, $p_T$~$>$~15~GeV for charged leptons and $\eta$~$<$~3, $p_T$~$>$~20~GeV for jets.
The acceptance cuts for the signal and background are taken to be $\eta$~$<$~5, $p_T$~$>$~15~GeV for positively charged leptons, $\eta$~$<$~3, $p_T$~$>$~20~GeV for jets and $\cancel{p}_T$~$>$~20~GeV for the missing transverse momentum.
% --*
The large pseudorapidity cut for the lepton is in anticipation of its forwardness and will be addressed shortly.
% *-- MODIFIED:
%Valid events are defined as having two jets with $\Delta R$~$>$~3.5, of which one is b-tagged, and one positively charged lepton (with taus excluded). 
Valid events are required to have exactly two jets with $\Delta R$~$>$~3.5, of which one is b-tagged, and one positively charged lepton (with taus excluded).
No restrictions are made on the number of negatively charged leptons.
% --*
At the partonic level, i.e.\ without considering realistic showering or detector simulations, the main backgrounds to this final state at the LHC are SM single top, $t\bar{t}$ and $W$ plus two jets.
A b-tagging efficiency of 50\% is assumed together with mistag rates of 0.5\%, 1.5\% and 10\% for the light quarks, gluons and charm quark respectively. 
To reduce the SM single top background the cut $p_T$~$<$~75~GeV is placed on the non-b-tagged jet.
This leaves $W$ plus two jets as the dominant background in the forward region.
% *-- MODIFIED:
%To reduce this background, a cut is placed on the reconstructed top mass: 150~GeV~$<m_t<$~190~GeV. 
To reduce this background, a cut is placed on the the top mass, 150~GeV~$<m_t<$~190~GeV, which is reconstructed from the four momenta of the charged lepton, b-tagged jet and neutrino. 
% --*
% *-- NEW:
The unknown neutrino momentum is deduced from the total missing transverse momentum, where the constraint $m_{l^+\nu}=m_W$ on the lepton-neutrino invariant mass is used to fix the longitudinal degree of freedom. 
% --*
% *-- NEW:
We have also examined the effect of extra jets on these backgrounds and found their impact to be marginal under the selected cuts. 
The largest effect is from $t\bar{t}$ plus one jet, which contributes an additional 10\% to the non-dominant $t\bar{t}$ background.
% --*

\begin{figure}[htbp]
  \centering
  \includegraphics[width=130mm]{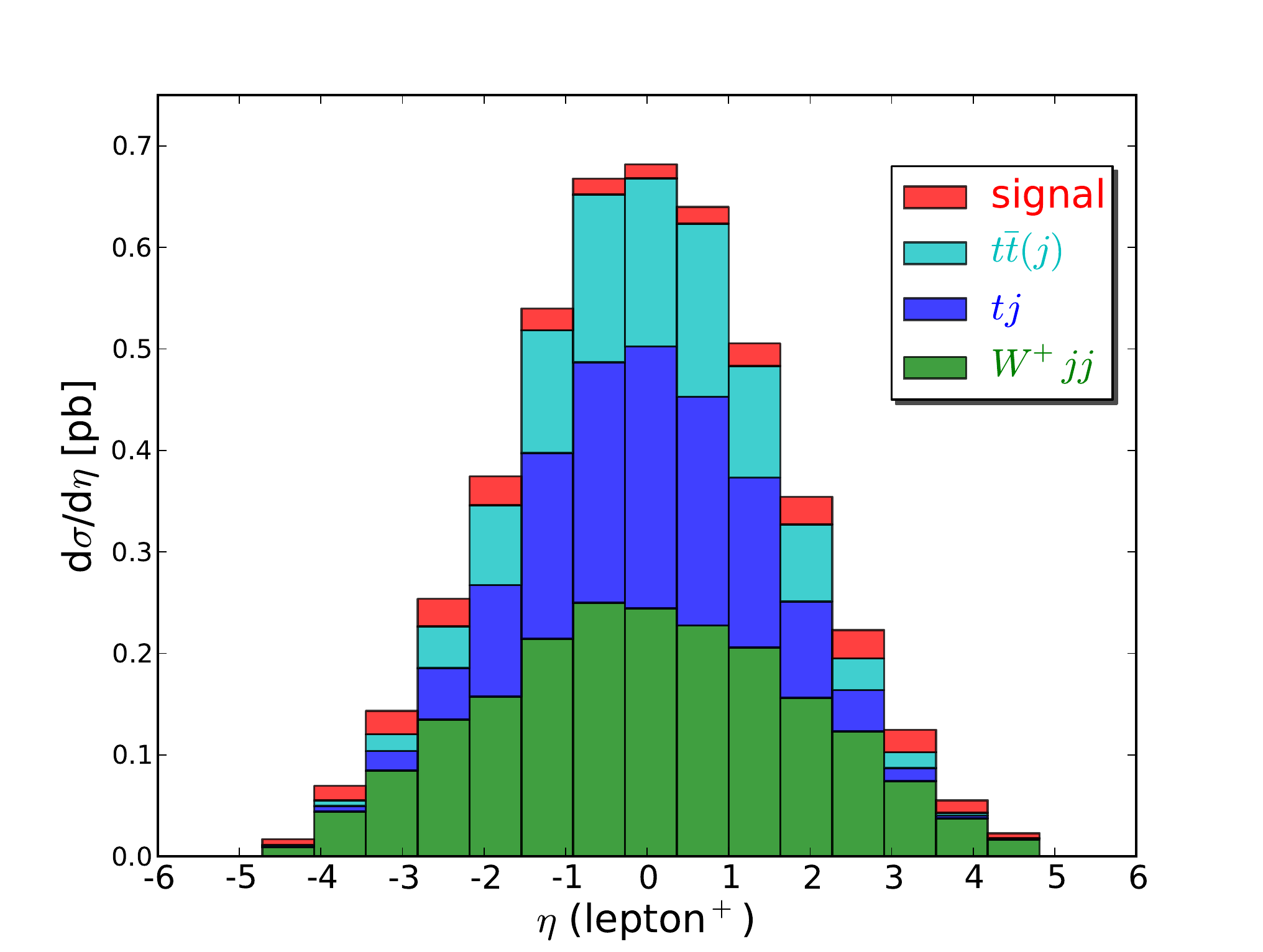}
  \caption{Pseudorapidity distribution of the final state lepton for the signal and SM background after cuts. As in in Figure~\ref{fig:forwardLepton}, the signal is seen to peak in the forward region. There is, therefore, a clear shape difference between the presence and absence of the signal when both are normalized against the central peak. \label{fig:signal}}
\end{figure}

The lepton pseudorapidity distribution of our prototype signal and its SM background, after cuts, is shown in Figure~\ref{fig:signal}.
As expected, the signal is most promising with respect to the background in the forward regions of an LHC detector.
Normalized against the central peak, there is a clear difference in shape between the presence and absence of the signal.
A minimum pseudorapidity cut of $\eta$~$\gtrsim$~1.5 on the lepton gives a reasonable signal over background ratio.
For instance, in the central bin the ratio is 2\% whereas in the bin at $\eta\sim 2$, where there is double the signal and almost half the background, the ratio is 8\%.
Of course, the upper bound of $\eta < 5$ for the lepton is only to illustrate its forwardness.
The LHC detectors ATLAS and CMS typically quote a pseudorapidity limit for muons between 2.5 and 3.
% *-- MODIFIED: REV2:
%Nonetheless, within this window of pseudorapidity the cross section remaining is still large enough to give ample statistical significance for an integrated luminosity of 30~fb$^{-1}$.
%With our selection criteria, and within the pseudorapidity window of 1.5~$<|\eta|<$~2.7, we find that an integrated luminosity of 10~fb$^{-1}$ gives a statistical significance of $S/\sqrt{B}$~=~9.6.
With our selection criteria, and within the pseudorapidity window of 1.5~$<|\eta|<$~2.7, the signal cross section is 100 fb and the $t\bar{t}(j)$, $tj$ and $Wjj$ background cross sections are 260, 320 and 600 fb respectively.  
An integrated luminosity of 10~fb$^{-1}$ therefore gives a statistical significance of $S/\sqrt{B}$~=~9.6.
% --*

Alternatively, our signal could be a candidate for the LHCb detector, which can measure very forward leptons and jets.
In this case, we would take 2~$<\eta<$~5 for our acceptance cuts together with $p_T$~$>$~25~GeV for jets and $p_T$~$>$~15~GeV for leptons.
Valid events are then defined as a charged lepton together with a b-tagged jet.
The catch is that we cannot adequately reduce the $t\bar{t}$ background in the region where the signal peaks, as LHCb is blind to the second top quark.
Beyond $\eta$~$\gtrsim$~3 the signal over background and significance are again adequate.
However, as the signal has already peaked, there is no longer a clear shape deviation.

The results given so far have been at the prototype parameter point. 
We now briefly discuss how the relevant parameters each affect the signal cross section.
A lighter gluino mass can significantly amplify the signal. Halving the gluino mass to $m_{\tilde{g}}$~=~1~TeV, for example, more than triples the cross section, whereas a heavier mass of $m_{\tilde{g}}$~=~3~TeV reduces it by 60\%.
We should keep in mind, however, that the mass of the gluino is related via experimental constraints to the level of flavour mixing allowed in the model. 
Decreasing the flavour mixing from its maximal value to $\theta_{LL}=\theta_{RR}=\pi/8$ has the effect of halving the cross section.
The cross section scales with the squark masses in the same way as it does with the gluino mass.
The size of the squark mass splitting, however, plays a crucial role for the cross section size due to the GIM-like mechanism present.
Decreasing the mass splitting to 200~GeV reduces the cross section to a tenth of the prototype's cross section.

% *-- MODIFIED: REV3:
We would like to reemphasize that the cross sections quoted in this paper were computed at leading order in QCD.
A further quantitative analysis would involve a full next-to-leading order analysis, which could have a significant impact on our results.
In particular, due to the forwardness of the signal and the presence of soft $p_T$ cuts, the presence of initial state radiation could seriously affect the signal significance.
% --*

% ### Conclusions
% ####################################################################
\section{Conclusions}
\label{sec:conclusions}

We have studied single top production at the LHC in a SUSY-QCD model with a heavy Dirac gluino.
The presence of a heavy Dirac gluino allows for less constrained top-up FCNCs. 
In this scenario, the FCNC process $ug\to tg$ was found to give the largest contribution.
This can be attributed to the up quark density in the colliding protons and t-channel gluon exchange in the leading diagram.
The key features of this signal are that the top quark is produced very forward and that it is asymmetric to its anti-top counterpart, as the latter lacks a valence quark.
The semileptonic decay of the top in turns gives a forward charged lepton whose pseudorapidity distribution is distinct in shape from that of SM single top production.

The signal was compared to its reducible SM background at the partonic level, without considering realistic showering or detector simulations.
The prototype parameter point chosen was found to have a promising signal over background ratio and statistical significance if a minimum pseudorapidity cut of 1.5 is placed on charged leptons.
Moreover, a difference in shape between the presence and absence of the signal is clearly visible in the forward regions of the lepton pseudorapidity distribution.
The feasibility of this signal is therefore dependent on the forward detection efficiency of muons at the LHC detectors.
If a shape difference for positively charged forward leptons can be detected, it would be suggestive of the top-up FCNCs we have examined in this paper.

% ### acknowledgements
% ####################################################################
\subsection*{Acknowledgments}

We would like to thank Ivo van Vulpen, Niels Tuning and Manouk Rijpstra for useful discussions.
This work was supported by the Netherlands Foundation for Fundamental Research of
Matter (FOM) and the National Organization for Scientific Research
(NWO). 

% ### bibliography
% ####################################################################
\bibliographystyle{utphys}
\bibliography{references}

\end{document}